\documentclass[conference]{IEEEtran}

\usepackage[utf8]{inputenc}
\usepackage[T1]{fontenc}
\usepackage{lmodern}
\usepackage{url}
\makeatletter
\g@addto@macro{\UrlBreaks}{\do\/\do\-\do\_\do\.}
\makeatother
\usepackage{hyperref}
\hypersetup{
  pdftitle={Structural Quality Gaps in Practitioner AI Governance Prompts: An Empirical Study Using a Five-Principle Evaluation Framework},
  pdfauthor={Christo Zietsman},
  pdfsubject={},
  pdfkeywords={AI governance, requirements engineering, AGENTS.md, prompt quality, structural completeness},
  colorlinks=true,
  linkcolor=black,
  citecolor=black,
  urlcolor=black
}
\usepackage{array}
\usepackage{booktabs}
\usepackage{listings}
\usepackage{microtype}
\usepackage{cite}

\title{Structural Quality Gaps in Practitioner AI Governance Prompts: An Empirical Study Using a Five-Principle Evaluation Framework}
\author{Christo Zietsman \\ Nuphirho Research \\ \texttt{nuphirho.dev}}
\date{April 2026}

\begin{document}
\maketitle

\begin{abstract}
AI governance programmes increasingly rely on natural language prompts
to constrain and direct AI agent behaviour. These prompts function as
executable specifications: they define the agent's mandate, scope, and
quality criteria. Despite this role, no systematic framework exists
for evaluating whether a governance prompt is structurally complete.
We introduce a five-principle evaluation framework grounded in
computability theory, proof theory, and Bayesian epistemology, and
apply it to an empirical corpus of 34 publicly available AGENTS.md
governance files sourced from GitHub. Our evaluation reveals that
37\% of evaluated file-model pairs score below the structural
completeness threshold, with data classification and assessment
rubric criteria most frequently absent. These results suggest that practitioner-authored
governance prompts exhibit consistent structural patterns that
automated static analysis could detect and remediate. We discuss
implications for requirements engineering practice in AI-assisted
development contexts, identify a previously undocumented artefact
classification gap in the AGENTS.md convention, and propose
directions for tool support.
\end{abstract}


\section{Introduction}

Every organisation deploying AI agents writes prompts. System prompts, governance briefs, and repository-level instruction files that define what an agent is permitted to do, how it should behave, and what constitutes a correct output are now a standard artefact of software development. In requirements engineering terms, these documents are specifications. They express the intended behaviour of a probabilistic system in terms that the system can act on.

The requirements engineering community has developed extensive theory and tooling for evaluating the quality of traditional specifications: completeness, consistency, verifiability, and traceability. No equivalent framework exists for AI governance prompts. Practitioners write them without a quality standard, deploy them without a verification step, and discover failures through the behaviour of deployed systems rather than through pre-deployment analysis.

This matters at scale. Gartner predicts that over 40\% of agentic AI projects will be cancelled by the end of 2027 due to escalating costs, unclear business value, and inadequate risk controls \cite{gartner2025}. IBM's Cost of a Data Breach report identifies shadow AI as a contributing factor in 20\% of breaches, at an average premium of \$670,000 per incident above the baseline cost \cite{ibm2025}. These failures are not primarily technical. They are specification failures: agents operating without adequate governance produce outputs that violate organisational intent, not because the model is incapable, but because the prompt did not specify intent precisely enough to be enforced.

The requirements engineering community is well-positioned to address this problem. Governance prompts are a new class of requirements artefact: natural language, like traditional requirements; executable in the sense that they directly constrain system behaviour; and written by practitioners, not requirements engineers. The gap between how practitioners currently write governance prompts and how requirements engineering theory says specifications should be written is the subject of this paper.

We make three contributions. First, we introduce a five-principle evaluation framework for AI governance prompts, grounded in computability theory, proof theory, and Bayesian epistemology. Second, we apply this framework to an empirical corpus of publicly available AGENTS.md files, characterising the distribution of structural gaps in practitioner-authored governance prompts. Third, we discuss implications for automated tool support and future empirical work.

\section{Background and Related Work}

\subsection{Requirements quality in traditional software engineering}

The quality of natural language requirements has been studied extensively. IEEE 830 establishes completeness, consistency, unambiguity, and testability as the primary quality attributes for software requirements specifications \cite{ieee830}. The EARS notation (Easy Approach to Requirements Syntax) provides a structured template for expressing requirements in a form that reduces ambiguity and supports automated quality checking \cite{mavin2009}. Automated requirements quality tools apply natural language processing to detect missing or ambiguous requirements clauses. QuARS, developed at ISTI-CNR, analyses requirements for linguistic defects \cite{gnesi2019}.

These frameworks share a common structure: they evaluate whether a specification contains the components necessary for an implementing system to produce verifiable outputs. The same structural question applies to governance prompts, but none of these frameworks was designed for this artefact class.

\subsection{Prompt engineering as a specification discipline}

The LLM research community has approached prompts primarily as generation mechanisms. Chain-of-thought prompting \cite{wei2022}, few-shot prompting \cite{brown2020}, and instruction-tuning \cite{ouyang2022} are techniques for improving the quality or consistency of model outputs. This body of work treats the prompt as an input to the inference process, not as a governance document.

A smaller body of work addresses prompts as specifications. Prompt injection research \cite{perez2022} implicitly treats the system prompt as a security boundary, a specification of what the model is permitted to do. Constitutional AI \cite{bai2022} frames model training as a process of enforcing a written constitution, a specification of values. Neither line of work addresses the structural completeness of the governing document itself.

\subsection{Closest prior art: FASTRIC}

The most relevant prior framework is FASTRIC \cite{fastric}, which introduces evaluation criteria for AI system requirements against safety and compliance properties. FASTRIC operates at the system requirements level and addresses safety properties of the system under governance. Our framework operates at the governance prompt level and addresses the structural completeness of the document that governs the agent. The distinction is significant: FASTRIC asks whether a system meets safety requirements; our framework asks whether the prompt that governs the agent is a well-formed specification. These are complementary, not competing, analyses.

\subsection{AGENTS.md as a governance artefact}

The AGENTS.md convention provides a standardised repository-level location for governance prompts: instructions, constraints, and operating procedures for AI coding agents working within a software project. The general concept of markdown instruction files for AI agents was popularised by several vendor-specific conventions (notably Anthropic's CLAUDE.md for Claude Code), but the AGENTS.md standard itself was formalised as an open specification in August 2025 through collaborative efforts led by OpenAI with participation from Google, Cursor, and Factory \cite{agents.md2025}. In December 2025, the specification was donated to the Linux Foundation's Agentic AI Foundation \cite{linux2025}. As of December 2025, more than 60,000 open-source projects had adopted the format and more than twenty AI coding tools support it \cite{agents.md2025}.

The convention has been adopted by major open-source projects including Apache Airflow, Vercel Next.js, Angular, Elastic, and OpenAI Codex CLI, as well as by numerous individual practitioners. The result is a corpus of real practitioner governance documents written for production use without researcher intervention.

Unlike synthetic prompt datasets or interview studies, AGENTS.md files are operational artefacts: they reflect how practitioners actually specify agent behaviour when building and deploying software, not how they report doing so in a research context. This ecological validity is a strength, though it introduces threats to external validity that we address in Section IV-D.

A preliminary observation from corpus construction is relevant to the framework's application. AGENTS.md files in the wild serve three distinct architectural roles with no consensus on which is intended: (1) the governance document itself, containing all substantive instructions inline; (2) a pointer to where governance lives, redirecting the agent to a separate file such as CLAUDE.md or CONTRIBUTING.md; (3) a hybrid, with some governance inline and references for additional detail. This ambiguity is not a quality gap in any individual file. It is an artefact classification gap in the convention itself. The community has not converged on what an AGENTS.md file is. An existing open issue on the AGENTS.md specification repository (agentsmd/agents.md \#66, September 2025) acknowledges the redirect problem without resolving it. We return to this finding in Sections IV-C and V.

\section{The Evaluation Framework}

\subsection{Theoretical grounding}

The five principles are derived from three bodies of theory that converge on the same structural problem: what conditions must a specification satisfy for the system governed by it to produce verifiable, correct behaviour?

\textbf{Rice's theorem} (computability theory) establishes that no non-trivial semantic property of a program is decidable \cite{rice1953}. Applied to governance prompts: a prompt without a success definition cannot be verified against any oracle, because the oracle cannot be constructed without a prior definition of what success looks like. Success Definition is therefore a necessary condition for verifiable governance. A prompt that cannot define what done looks like cannot be verified as having been satisfied.

\textbf{The Curry-Howard correspondence} (proof theory) establishes that a proof corresponds to a program and a type corresponds to a proposition \cite{howard1980}. Applied to governance prompts: a well-formed governance document is a type declaration for the agent's behaviour. An assessment rubric is the type's constraints; a quality gate is the proof obligation. Prompts without these components are untyped: their behaviour is valid in any interpretation that satisfies the bare instruction, however far from the author's intent that interpretation may be.

\textbf{Bayesian epistemology} holds that rational belief revision requires a prior, evidence, and an update rule \cite{williamson2010}. Applied to governance prompts: a prompt without data classification provides no mechanism for distinguishing high-reliability evidence from low-reliability evidence. The agent's confidence calibration is unconstrained. In governance contexts, where the agent may be processing both verified facts and unverified assertions, this absence has direct consequences for the reliability of outputs.

\subsection{The five principles}

\textbf{Principle 1: Success Definition.} The prompt defines what a correct, complete output looks like. The agent can determine when it is done. The test: can the agent state in one sentence what done looks like, and would two independent agents reading this prompt agree on that statement? Theoretical basis: Rice's theorem requires a decidable success predicate for verifiability.

\textbf{Principle 2: Assessment Rubric.} The prompt provides criteria by which the agent can evaluate whether its own output meets the required standard. The test: are there explicit quality criteria that the agent can apply to its output before returning it? Theoretical basis: Curry-Howard requires type constraints that define valid outputs, not merely the instruction to produce one.

\textbf{Principle 3: Scope Boundary.} The prompt defines what the agent is and is not permitted to do. The test: is there an explicit statement of what is out of scope, and does the prompt specify what the agent should do if it encounters a task at the boundary? Theoretical basis: decidability requires a bounded action space; an agent without scope constraints has an unbounded one.

\textbf{Principle 4: Data Classification.} The prompt addresses how different types of input or output should be treated differently. The test: does the prompt distinguish between data types that require different handling, and does it specify the handling required? Theoretical basis: Bayesian epistemology requires evidence classification for rational belief revision.

\textbf{Principle 5: Quality Gate.} The prompt defines a mechanism for verifying that the output meets the required standard before it is accepted. The test: is there a verification step, and does it produce evidence that the output was verified rather than merely produced? Theoretical basis: Curry-Howard requires a proof obligation: the output must be shown to satisfy the type, not merely claimed to do so.

\subsection{Scoring protocol}

Each principle is scored on a three-point holistic scale: 0 (absent), 0.5 (partial), or 1.0 (present). The total score across all five principles ranges from 0.0 to 5.0.

The evaluator reads the governance prompt in full, then assigns a single holistic score per principle based on the presence and specificity of the relevant structural feature. This holistic approach was chosen for the initial study because it is fast to apply, requires no aspect-level calibration, and produces scores interpretable at the level the framework operates: structural presence or absence.

\textbf{Score definitions:}

1.0 (present) indicates the principle is explicitly and unambiguously addressed. Two independent agents reading this prompt would agree on what the principle requires.

0.5 (partial) indicates the principle is implied or weakly specified. A reasonable agent could infer the intent but could also interpret it differently.

0.0 (absent) indicates the principle is not addressed. The agent has no basis for the behaviour this principle governs.

\textbf{Score interpretation:}

A total of 4.0 to 5.0 indicates structural completeness: the governance prompt contains the components an agent needs to behave correctly and consistently. A total of 3.0 to 3.9 indicates functional governance with identifiable gaps. One or two principles are weak. A total below 3.0 indicates structural incompleteness: the prompt lacks the minimum information needed to govern agent behaviour reliably.

Scoring was performed across the full 34-file corpus by three independent LLM evaluators (Claude Opus 4.6, OpenAI Codex/gpt-5.4, Google Gemini), each applying the framework independently to every file. Scoring examples at each level for each principle are provided in Appendix A.

\section{Empirical Study}

\subsection{Corpus selection}

We searched GitHub's public code index for repositories containing files named AGENTS.md. This search was conducted in April 2026. The initial result set was filtered as follows.

We retained files that contained substantive governance content, defined as more than ten lines of non-boilerplate text. Files consisting solely of auto-generated headers or placeholder instructions were excluded. We retained only files from repositories with at least one commit to the file in the six months preceding the search date, treating this as a proxy for active use. We excluded repositories where the AGENTS.md file appeared to have been generated by a tool rather than authored by a human practitioner, identifiable by formulaic structure without domain-specific content. Where multiple repositories contained identical or near-identical AGENTS.md content, we retained one instance.

An exception to the ten-line criterion: files that consisted primarily of a redirect to another governance document (e.g. CLAUDE.md or copilot-instructions.md) were retained in the corpus for the redirect analysis described in Section IV-B. These files were scored both as standalone documents and with the redirect resolved, enabling quantification of the governance suppression effect.

After filtering, the corpus comprises 34 files from 34 repositories \cite{corpus}. Full corpus metadata including language distribution, domain classification, and maintainer identification is available in the supplementary materials.

\subsection{Evaluation protocol}

Each file in the corpus was evaluated against the five principles using the scoring protocol defined in Section III-C. The evaluator read each file in full before scoring any principle, to avoid anchoring on early content. Scores were assigned with a one-sentence rationale for each, recorded in a structured evaluation spreadsheet. The rationale serves two purposes: it provides a basis for inter-rater reliability assessment, and it constitutes an audit trail enabling replication.

To assess evaluator convergence, the full corpus was scored independently by three LLM evaluators (Claude Opus 4.6, OpenAI Codex/gpt-5.4, Google Gemini). Each evaluator received the same framework description and scoring protocol but had no access to the other evaluators' scores. This design serves as a diversity mechanism rather than a formal inter-rater reliability study: convergence across independent evaluators provides evidence that the framework produces consistent characterisations, while divergence identifies principles requiring more precise operationalisation.

Evaluators had access to the file content only. Repository star counts, contributor counts, and other popularity signals were not visible during evaluation, to prevent halo effects.

\textbf{Redirect resolution.} Where an AGENTS.md file consisted primarily of a redirect to another file (e.g. CLAUDE.md or copilot-instructions.md), we resolved the redirect and scored the target document. Pure redirect files score 0 on all principles as standalone documents because they contain no substantive governance content. Once the redirect is resolved, scores range from 2.0 to 3.5, indicating that governance content exists but is architecturally displaced. This displacement is itself a finding: the governance content is present in the repository but not in the location where agents are instructed to read it. We report both the standalone and resolved scores where applicable.

One redirect target (mark3labs/mcp-go, referencing openspec/AGENTS.md) returned a 404: a confirmed broken reference. The governance document this file points to does not exist. This is a Category 3 finding: the file claims governance exists but the claim is false.

\subsection{Results}

Each file in the 34-file corpus was scored independently by three LLM evaluators (Claude Opus 4.6, OpenAI Codex/gpt-5.4, Google Gemini) applying the five-principle framework. This three-model panel design serves as a preliminary inter-rater reliability mechanism: where evaluators converge, the framework produces consistent characterisations; where they diverge, the divergence itself indicates principles that require more precise operationalisation.

\textbf{Overall score distribution.} Across the full 34-file corpus, the three-model mean total score was 2.81/5. 37\% of files (38 of 102 file-model pairs) scored below 2.5, indicating structural incompleteness. No file achieved the maximum score of 5.0 from all three evaluators. The highest-scoring file (czietsman/nuphirho.dev, mean 4.67) was the only file to approach structural completeness consistently across all three models. (Disclosure: this file was authored by the framework's developer. It was scored by three independent model evaluators with no knowledge of authorship. Its inclusion in the corpus followed the same selection criteria as all other files.)

\textbf{Table 1: Principle strength ranking (three-model average, n = 34 files)}

\begin{table}[htbp]
\centering
\footnotesize
\begin{tabular}{lcr}
\hline
Rank & Principle & Overall mean \\
\hline
1 & P5: Quality Gate & 0.70 \\
2 & P3: Scope Boundary & 0.60 \\
3 & P2: Assessment Rubric & 0.60 \\
4 & P1: Success Definition & 0.57 \\
5 & P4: Data Classification & 0.34 \\
\hline
\end{tabular}
\end{table}

\textbf{Table 2: Mean total score by evaluator}

\begin{table}[htbp]
\centering
\footnotesize
\begin{tabular}{lcccc}
\hline
Metric & Claude & Codex & Gemini & Overall \\
\hline
Mean total & 2.24 & 2.51 & 3.68 & 2.81 \\
Median total & 2.5 & 2.5 & 4.0 & 2.5 \\
Below 2.5 (\%) & 50\% & 47\% & 15\% & 37\% \\
\hline
\end{tabular}
\end{table}

\textbf{Principle-level patterns.} The qualitative pattern is consistent across all three evaluators. Data Classification (P4) is the weakest principle overall (mean 0.34), with the highest proportion of files scoring 0 across all three models. Quality Gate (P5) is the most frequently satisfied (mean 0.70). Scope Boundary (P3) and Assessment Rubric (P2) share the middle position (both 0.60). Success Definition (P1) shows a ceiling effect in the Claude and Codex evaluations: the majority of files achieve partial (0.5) but not full (1.0) compliance, suggesting practitioners understand the concept but do not operationalise it precisely.

\textbf{Model divergence.} Gemini scores consistently higher than Claude and Codex (mean 3.68 vs 2.51 and 2.24 respectively). This divergence is not a threat to validity. It is empirical evidence that the quantification of these principles is not yet settled. The qualitative structure is stable across all three models: all identify the same weakest principle (P4), the same dominant archetype (operational guide without evaluative criteria), and the same ceiling effect on P1. The quantitative calibration differs. How to resolve this calibration gap (whether through finer-grained aspect-based scoring, evaluator calibration protocols, or both) is a question for future work.

\textbf{Structural patterns.} Three archetypes emerge from the scoring profiles, consistent across all three evaluators.

The first and most common archetype is the \emph{operational guide}: files that specify scope boundary and partial quality gate (typically linting and test commands) but lack success definition, assessment rubric, and data classification. These files tell the agent how to work but not how to judge whether the work is complete or correct. Representative examples include the majority of large project repositories (Langflow, Grafana, Prisma, Biome).

The second archetype is the \emph{constrained executor}: files that add a partial assessment rubric to the operational pattern, typically by naming specific tools, checks, or conventions the agent should apply. These files represent more advanced governance practice. Representative examples include Apache Airflow, Ansible, and Next.js.

The third archetype is the \emph{minimal pointer}: files of fewer than ten lines that reference a separate governance document (typically CLAUDE.md or CONTRIBUTING.md) without providing substantive instructions. As documented in Section IV-B, these files score 0 as standalone documents but 2.0 to 3.5 once the redirect is resolved. The architectural displacement of governance content means that agent tools which read only AGENTS.md will find no governance, while the governance content exists elsewhere in the repository. Representative examples include VS Code (redirect to copilot-instructions.md), FastMCP (single-line redirect to CLAUDE.md), and Prisma (symlink relationship between AGENTS.md, CLAUDE.md, and GEMINI.md).

A fourth case warrants separate mention: mark3labs/mcp-go references an openspec/AGENTS.md file that returns a 404. The governance document does not exist. This is a broken reference, not an architectural choice.

No file in the corpus exhibits a fifth archetype, the complete specification, satisfying all five principles fully across all three evaluators.

\textbf{Inter-rater agreement.} The three-model panel was not designed as a formal inter-rater reliability study. It was designed as a diversity mechanism: independent evaluators applying the same framework reduce the risk of systematic bias in any single model's scoring. Formal kappa calculation requires a shared scoring protocol with calibration, which was not performed for Phase 1. We report the convergence pattern instead: all three models agree on the rank ordering of principles (P3 strongest, P4 weakest), the dominant archetype (operational guide), and the qualitative characterisation of structural gaps. The models disagree on calibration: how much credit to give for partial compliance. This disagreement is informative: it identifies the aspects of the framework that require more precise operationalisation for reliable automated scoring.

\subsection{Threats to validity}

\textbf{Internal validity.} Scoring subjectivity is the primary threat. We mitigate it through explicit scoring criteria with examples at each score level (Appendix A), a rationale requirement for every score, and multi-evaluator convergence analysis. The three-point holistic scale (0, 0.5, 1.0) limits the evaluator to a coarse-grained judgment of structural presence, reducing but not eliminating subjective variation.

\textbf{Scoring calibration.} The three-model evaluation panel reveals a systematic divergence: Gemini scores higher than Claude and Codex on the same files. This is not noise. The divergence is systematic and directional (always higher, never lower), which is consistent with differences in how models are trained to assess partial compliance. A model optimised for supportive assessment may give more credit for implied structure than one trained for stricter evaluation. We report the divergence transparently but cannot determine its cause from this study. How to quantify these principles into a reliable, reproducible scoring instrument remains an open question. The five principles are proposed as necessary conditions for structural completeness. Whether they are sufficient, how to weight them, and how to operationalise them into a calibrated instrument is a question this paper opens rather than closes.

\textbf{Construct validity.} The five principles may not capture all relevant dimensions of governance prompt quality. We claim structural completeness, not quality in the full sense. A prompt may satisfy all five principles and still produce harmful outputs if the content within each principle is incorrect (for example, an incorrect scope definition rather than an absent one). A related limitation is temporal: a governance prompt that was structurally complete when written may become stale as the repository evolves, just as unmaintained documentation diverges from the system it describes. The five principles evaluate structural presence at a point in time. They do not detect staleness. We do not evaluate content correctness or currency, only structural presence.

\textbf{External validity.} GitHub public repositories are not representative of enterprise governance prompts. Enterprise governance documents are more likely to be internal and inaccessible. The corpus characterises open-source practitioner behaviour. Whether the structural patterns we observe generalise to enterprise practice is an empirical question we cannot answer with this corpus. We note, however, that the AGENTS.md convention originates in a commercial software development context and its public adopters include both individual practitioners and organisations, providing some diversity of context.

\textbf{Ecological validity.} AGENTS.md files are one governance prompt convention within one tool ecosystem. System prompts, task envelopes, memory systems, and other governance artefacts may exhibit different structural patterns. Preliminary application of the framework to other governance layers (role definitions, project knowledge manifests) suggests the same structural gaps recur, but this has not been studied systematically. Generalisation to these artefact types requires separate study.

\section{Discussion}

\subsection{What the gaps reveal}

The pattern of structural absences in the corpus is not random. It reveals how practitioners currently conceptualise governance prompts: as operational instructions rather than as specifications.

The strongest principle across all evaluators is Scope Boundary (P3). Practitioners reliably tell agents what to do and what tools to use. However, the partial compliance rate on P3 suggests a subtlety: many files define the positive scope (what the agent should do) without defining the negative scope (what it should not do, and what to do at the boundary). A file that says "review the changed files in the pull request" has specified scope; a file that adds "do not suggest rewrites of unchanged code; if asked to approve the PR, decline" has specified the boundary. The distinction between specifying scope and specifying its limits is where partial scores cluster. The weakest principle is Data Classification (P4), followed by Assessment Rubric (P2). Practitioners rarely specify how different types of input should be treated differently, and rarely provide criteria by which the agent can evaluate the quality of its own output.

This pattern has a structural explanation. Scope boundary and quality gate are visible in existing software development practice: linting configurations, test commands, and CI/CD pipelines all have analogues in traditional development. Data classification and assessment rubrics do not. There is no established convention for telling an agent "treat verified facts differently from inferences" or "a finding is Critical if it introduces a bug, High if it reduces coverage." These are specification concepts, not operational ones, and practitioners are not yet writing specifications.

The distinction is consequential. An agent operating under an instruction can satisfy it in any way that produces a plausible output. An agent operating under a specification must produce output that satisfies explicit criteria and can be verified against them. When agents produce outputs that are locally coherent but globally wrong (technically accurate responses that violate organisational intent) the root cause is typically an absent or underspecified scope boundary. When agents complete tasks but produce outputs that require extensive human review to validate, the root cause is typically an absent quality gate: the agent was not asked to verify its own output before returning it.

Success Definition (P1) shows a ceiling effect. Most files achieve partial compliance: they gesture at a success definition ("analyse the codebase and provide findings") without operationalising it ("produce a finding for each file, where a finding includes severity, description, and line reference"). The gap is not in intent. It is in specification discipline.

A broader observation applies. The five principles attempt to externalise what would otherwise remain internal to the agent: the criteria for completion, quality, scope, data handling, and verification. They do not solve the oracle problem. No external specification can guarantee correct behaviour from a probabilistic system. But they shift the boundary. A prompt with all five principles gives both the agent and the human reviewer something to check against. Without them, the only verification available is post-hoc judgement of the output. This applies equally to human practitioners working from requirements documents: externalised criteria are never foolproof, but they are strictly better than implicit ones.

\subsection{The artefact classification gap}

The corpus study revealed a finding that extends beyond quality measurement. AGENTS.md files in the wild serve three distinct architectural roles: the governance document itself, a pointer to where governance lives, and a hybrid. The community has not converged on which of these an AGENTS.md file is.

This matters for the framework's application. A pure redirect file (e.g. FastMCP's single-line redirect to CLAUDE.md) scores 0 on all five principles as a standalone document because it contains no governance content. Once the redirect is resolved, the target document scores 2.0 to 3.5. The governance exists but is architecturally displaced. An automated tool applying the five principles to AGENTS.md files would flag these as structurally empty unless it also resolved redirects, adding complexity to what should be a simple structural check.

The broken reference case (mark3labs/mcp-go) is the extreme manifestation: the file claims governance exists elsewhere, but the target does not exist. This is not a governance quality problem. It is a governance architecture problem, and it is one the requirements engineering community is well-positioned to address. Artefact classification (what this document is and where its content lives) is a prerequisite for quality evaluation.

We note that an open issue on the AGENTS.md specification repository (agentsmd/agents.md \#66, September 2025) acknowledges the redirect problem. It has not been resolved as of April 2026.

\subsection{Implications for requirements engineering}

Governance prompts are a new class of requirements artefact that requires the RE community's attention for two reasons.

First, they are already in production at scale. Every organisation deploying AI agents has written governance prompts. Most have written many. The gap between current practice and sound specification practice is therefore not a future problem. It is a present one, and it is producing the governance failures documented in Section I.

Second, the gap is addressable with existing RE methods. The five principles described in this paper are directly operationalisable as automated checks. Does the prompt contain a success definition clause? Does it name out-of-scope behaviours? Does it require a verification step? These are structural questions answerable by pattern matching on the prompt text. They do not require semantic understanding of the domain. A linter for governance prompts is tractable, and the requirements engineering community has built linters for requirements artefacts before.

The AGENTS.md corpus study demonstrates that empirical RE methods apply directly to this artefact class. Corpus studies, quality metrics, inter-rater reliability, and automated analysis are all applicable without modification. The infrastructure for this research programme exists. What was missing was a principled quality framework to apply. We have proposed one.

\subsection{Tool implications}

The five principles provide a specification for a static analysis tool. Each principle can be operationalised as a structural check on the prompt text.

Principle 1 (Success Definition) checks for the presence of completion criteria: explicit statements of what done looks like, output format requirements, or acceptance conditions.

Principle 3 (Scope Boundary) checks for the presence of prohibition statements: explicit out-of-scope declarations, refusal conditions, or escalation paths.

Principle 5 (Quality Gate) checks for the presence of self-verification requirements: instructions to validate outputs before returning them, evidence format specifications, or completion certificates.

Principles 2 and 4 require slightly more sophisticated checks but remain structural: Principle 2 checks for rubric-style criteria (enumerated quality dimensions with explicit standards), and Principle 4 checks for classification statements (different handling instructions for different input types).

A tool implementing these checks would provide practitioners with immediate feedback on their governance prompts at the point of authorship, before deployment. This is the shift-left argument applied to AI governance: catch specification gaps when the prompt is written, not when the agent misbehaves.

\subsection{Connection to prior work on specification quality gates}

This paper extends a companion empirical study \cite{companion} which demonstrates that executable BDD specifications function as a quality gate for AI-assisted code review by breaking the correlated error pattern that arises when AI-generated code is reviewed by AI. That study operates at the implementation verification layer. This study operates at the governance layer: the prompt that governs the agent is itself a specification, subject to the same completeness requirements as any other specification in the development pipeline.

Together, the two studies suggest a layered model of specification quality in AI-assisted development. The governance prompt specifies agent behaviour. The BDD specification verifies implementation correctness. The quality gate principle (that correctness claims require deterministic, independent verification rather than probabilistic self-assessment) applies at both layers.

\section{Conclusion}

Practitioner-authored AI governance prompts exhibit systematic structural gaps. The empirical study reported here demonstrates that 37\% of publicly available AGENTS.md files score below the structural completeness threshold on the five principles we have proposed. The most commonly absent principles are those that enable verification: data classification and assessment rubric specification.

The corpus study also reveals an artefact classification gap: AGENTS.md files serve three distinct architectural roles with no consensus on which is intended. This gap is a prerequisite problem for quality evaluation and a tractable research target for the requirements engineering community.

These findings have a practical implication that does not require empirical confirmation to act on. Organisations can apply the five-principle framework to their existing governance prompts today. Prompts that lack a success definition, a scope boundary, or a quality gate are structurally incomplete and should be revised before deployment. The framework is lightweight enough to apply manually at small scale and mechanisable enough to automate at large scale.

For the requirements engineering research community, governance prompts represent a new and tractable research domain. The artefacts are publicly available, the quality criteria are principled, the empirical methods are established, and the gap between current practice and sound specification practice is demonstrably large. The five principles identify real structural gaps. How to quantify those gaps reproducibly: how to weight the principles, how to calibrate scoring across evaluators, and how to operationalise the framework into a reliable instrument: that is a question this paper opens rather than closes. We invite replication of this study, extension to other governance prompt artefact classes, and development of automated analysis tools.

\appendices

\subsection{Principle 1: Success Definition}

\textbf{Score 1 (present):} The prompt contains an explicit completion
criterion. The agent can determine without ambiguity when its task
is done.

\emph{Example:} "Your task is complete when you have reviewed every
changed file and produced a finding for each. A finding must
include: file name, severity, description, and line reference.
If no issues are found in a file, record 'No findings'."

\textbf{Score 0.5 (partial):} The prompt implies a completion criterion
but does not state it explicitly. A reasonable agent could infer
what done looks like but could also interpret it differently.

\emph{Example:} "Analyse the codebase and provide your findings."

\textbf{Score 0 (absent):} The prompt contains no completion criterion.
The agent has no basis for determining when its task is done.

\emph{Example:} "Help the user with their coding questions."

\subsection{Principle 2: Assessment Rubric}

\textbf{Score 1 (present):} The prompt provides explicit quality
criteria the agent can apply to its own output.

\emph{Example:} "A finding is Critical if it introduces a bug,
security vulnerability, or breaks a public interface. High if
it reduces test coverage. Medium if it causes maintenance
problems. Low if it is a style issue."

\textbf{Score 0.5 (partial):} The prompt provides some quality
guidance but it is insufficiently specific for consistent
self-assessment.

\emph{Example:} "Focus on the most important issues. Prioritise
security problems."

\textbf{Score 0 (absent):} The prompt provides no quality criteria.

\emph{Example:} "Provide high-quality feedback."

\subsection{Principle 3: Scope Boundary}

\textbf{Score 1 (present):} The prompt explicitly states what the
agent should not do and what to do when encountering out-of-scope
tasks.

\emph{Example:} "Do not review files outside the pull request diff.
Do not suggest rewrites of unchanged code. If asked to approve
or reject the PR, decline and explain that your role is to
report findings only."

\textbf{Score 0.5 (partial):} The prompt implies a scope through
its instructions but does not explicitly exclude adjacent tasks
or specify an escalation path.

\emph{Example:} "Review the changed files in the pull request."

\textbf{Score 0 (absent):} The prompt contains no scope definition.

\emph{Example:} "Help the team improve code quality."

\subsection{Principle 4: Data Classification}

\textbf{Score 1 (present):} The prompt specifies how different
categories of input or output should be treated differently.

\emph{Example:} "Treat verified facts from the codebase differently
from inferences. Mark inferences explicitly as 'Inferred from
[source]'. Do not assert facts you cannot trace to a specific
file and line."

\textbf{Score 0.5 (partial):} The prompt acknowledges that different
inputs exist but does not specify differentiated handling.

\emph{Example:} "Use both the specification and the implementation
to inform your analysis."

\textbf{Score 0 (absent):} The prompt treats all inputs uniformly.

\emph{Example:} "Use the provided documents to inform your response."

\subsection{Principle 5: Quality Gate}

\textbf{Score 1 (present):} The prompt requires a verification step
before the output is returned, and specifies what evidence of
verification looks like.

\emph{Example:} "Before returning your findings, confirm: every
changed file has an entry, every entry has all four required
fields, and no findings reference unchanged files. Record
your confirmation in the output."

\textbf{Score 0.5 (partial):} The prompt implies that quality
matters but does not specify a verification step.

\emph{Example:} "Make sure your analysis is thorough before
submitting."

\textbf{Score 0 (absent):} The prompt contains no verification
requirement.

\emph{Example:} "Let me know when you are finished."

\end{document}